# Coherent microscopy at resolution beyond diffraction limit using post-experimental data extrapolation

Tatiana Latychevskaia* and Hans-Werner Fink

Physik Institut der Universität Zürich

Winterthurerstrasse 190

CH- 8057 Zürich

Switzerland

**Abstract**

Conventional microscopic records represent intensity distributions whereby local sample information is mapped onto local information at the detector. In coherent microscopy, the superposition principle of waves holds; field amplitudes are added, not intensities. This non-local representation is spread out in space and interference information combined with wave continuity allows extrapolation beyond the actual detected data. Established resolution criteria are thus circumvented and hidden object details can retrospectively be recovered from just a fraction of an interference pattern.

*Corresponding author: tatiana@physik.uzh.ch

**Main text**

It is a generally accepted notion that once a microscopy experiment has been carried out, the resolution of the experimental record is an intrinsic property of the various experimental parameters and thus fixed once the experiment has been completed.

Ever since Ernst Karl Abbe introduced the term "Numerical Aperture" (N.A.) and proposed the resolution criterion $R=\lambda/(2\text{N.A.})$[1,2], it has been the quantitative measure of optical system performance until today. However, with the invention of optical lasers, and later, coherent X-ray and electron sources, imaging techniques employing coherent waves have been developed, and here the Abbe's criterion is only remotely related to the possibly achievable resolution. Coherent radiation, despite of many obvious advantages, deteriorates the resolution due to interference effects between the scattered waves. For example, for two point scatterers, the total intensity in case of incoherent radiation is given by: $I = |U_1|^2 + |U_2|^2$ where $U_1$ and $U_2$ are complex-valued waves diffracted by scatterers 1 and 2 while in case of coherent radiation, the total intensity is given by $I = |U_1|^2 + |U_2|^2 + U_1U_2* + U_1*U_2$. Now, the interference term $U_1U_2* + U_1*U_2$ obscures the image of two scatterers. However, this very interference term contains the phase information about the interfering waves, and, in the technique we propose here, it allows reconstructing the entire complex-valued wavefront created by the scatterers.

Previously, it has been reported that provided the complex-valued scattered wavefront, in particularly its phase, is known, it can be extrapolated beyond the size of the recorded interference pattern increasing the resolution of the reconstructed object[3]. The ingenious way of providing such phase information is holography, where the unknown object wave is superimposed with a well-known reference wave[4,5]. However, in a general case of coherent imaging, a reference wave is not provided and only the amplitude of the complex-valued scattered wave can be captured, thus the method[3] cannot be applied. Here we propose a universal approach for post-extrapolation of experimental coherent interference patterns that allows extrapolation and resolution enhancement even without phase information available from an experimental record.

In a typical experiment, a finite fraction of an interference pattern $I_0$, such as a hologram or coherent diffraction pattern, is recorded by a detector of size $S_0 \times S_0$, and digitized with $N_0 \times N_0$ pixels, so that $S_0 = N_0\Delta$, where $\Delta$ is the pixel size of the detector. The complex-valued wave $U_0$ forming the interference pattern $I_0$ can be reconstructed by employing conventional numerical methods. The back-propagation of the wave $U_0$ to the object domain results in the reconstruction of the object at a resolution provided by the Abbe criterion $R_0 = \lambda/(2\text{N.A.}_0)$, where N.A.$_0$ is limited by the detector size $S_0$. The key of our method is that the distribution $U_0$ is complex-valued and thus contains sufficient information to uniquely define the elementary waves scattered by the object. These elementary waves can be extrapolated well beyond the detector of size $S_0 \times S_0$, and thus effectively increase the numerical aperture and hence the resolution. Thus, obtaining the complex-valued distribution of $U_0$ constitutes the first step of our method. The second step consists of extrapolation of $U_0$. Here an iterative routine is applied, which includes the following steps:

(i) Formation of the input of the complex-valued field in the detector plane $U(x_s,y_s)$. For the first iteration, the reconstructed complex-valued distribution $U_0$ of the size $\Delta N_0 \times \Delta N_0$ (or $S_0 \times S_0$) is extended onto an area $\Delta N \times \Delta N$ (or $S \times S$) by padding $U_0$ with random complex-valued numbers. Here $N > N_0$ (or $S > S_0$) while the pixel size $\Delta$ remains unchanged. The amplitude of the central $S_0 \times S_0$ part (sampled with $N_0 \times N_0$ pixels) is replaced by the square root of the measured intensity $\left| U_0(x_s,y_s) \right|$ after each iteration. Following each iteration, the amplitude of the remaining $S \times S$ part (sampled with $N \times N$ pixels) and the phase distribution of the entire field are updated.

(ii) Back propagation to the object plane. In the case of coherent diffraction imaging, it is calculated by a backward Fourier transform. In the case of holography, the integral transformation is given by the Fresnel-Huygens principle and must be computed[6-7].

(iii) In the object plane, the following constraints are applied to the reconstructed complex-valued object distribution $o(x_o,y_o)$. Since the object exhibits a finite size, the distribution $o(x_o,y_o)$ is multiplied with a loose mask and the values outside the mask are set to zero[8]. Additional constraints, such as a real, positive and finite transmission function of the object or non-negative absorption (the latter we used in the presented here work), may also be applied[9]. This results in an updated transmission function $o'(x_o,y_o)$.

(iv) The updated complex-valued wavefront in the screen plane $U'(x_s,y_s)$ is obtained by forward propagation and its amplitude and phase distributions are the input values for the next iteration starting at step (i).

The initially random numbers are updated after each iteration and eventually turn into an extrapolated interference pattern beyond the experimental record. The resulting self-extrapolated interference pattern of size $S \times S$ provides a nominally larger numerical aperture $NA > NA_0$. As a consequence, a resolution better than $R_0$, respectively beyond the diffraction limit is achieved for the reconstructed object.

To demonstrate our method we select the most popular modern coherent imaging techniques - coherent diffraction imaging (CDI)[10], which is applied to single-particle diffraction patterns recorded at free electron laser facilities[11-12]. CDI allows a complete recovery of a non-periodic object from its far-field diffraction pattern, provided the latter is sampled with at least twice the Nyquist frequency (oversampling)[13], by using one of the iterative phase retrieval routines[14-17]. These routines are based on propagation of the optical field between detector and object plane, calculated by Fourier transforms. Since numerical Fourier-transformations are performed on finite sized images, this automatically imposes another constraint: the experimental diffraction pattern is surrounded by zeros, while in an idealized experiment it is not. By applying our technique we just remove the constraint of the diffraction pattern being zero-padded and instead let the experimental diffraction pattern extrapolate itself.

A simulated interference pattern created by two coherent point sources is shown in Fig. 1. Poisson distributed noise was added to the simulated diffraction pattern to mimic a realistic experiment with photons or electrons. The intensity distribution in the far-field resembles an equidistant fringes pattern, as shown in Fig. 1(a). When this far-field diffraction pattern is recorded in the oversampling regime, the two scatterers can be recovered by using one of the iterative phase retrieval methods[14], the result is shown in Fig. 1(b). When just a fraction of the interference pattern $I_0$ (of size $S_0 \times S_0$ sampled with $200 \times 200$ pixels, as marked with the red square in Fig. 1(a)) is available, as shown in Fig.1(c), the related complex-valued wave distribution at the detector $U_0$ is retrieved with a phase retrieval procedure, but the two reconstructed point sources are barely resolved, see Fig. 1(d). Here, the poor resolution of the reconstruction is given by the limited extend $S_0 \times S_0$ of the diffraction pattern $I_0$. Next, provided the phase distribution of $U_0$ has been recovered, we extrapolate complex-valued $U_0$ beyond $S_0 \times S_0$ ($200 \times 200$ pixels) area up to $S \times S$ ($1000 \times 1000$ pixels) as described above. In the first iteration we pad the outside region of $U_0$ up to $1000 \times 1000$ pixels with random numbers, as depicted in Fig. 1(e), and allow for updating the values of these pixels after each iterative run. For extrapolation, a slow-convergent iterative algorithm, such as error-reduction algorithm[14] is preferred, as it provides stable continuous decrease of error function $\mathrm{Error} = \dfrac{\sum\limits_{i,j} \left| |U| - |U_0| \right|}{\sum\limits_{i,j} |U|}$. After 10000 iterations, the interference pattern extrapolated itself noise-free beyond the area of the actual data of $I_0$, see Fig. 1(g) and Fig.2. As a result, the N.A. has effectively been increased and the reconstruction of the self-extrapolated diffraction pattern demonstrates superior resolution; the two point sources are now clearly resolved, as shown in Fig. 1(h). Thus, from just a fraction of a diffraction pattern it is possible to extrapolate the entire interference pattern beyond the recorded area and, eventually, enhance the resolution.

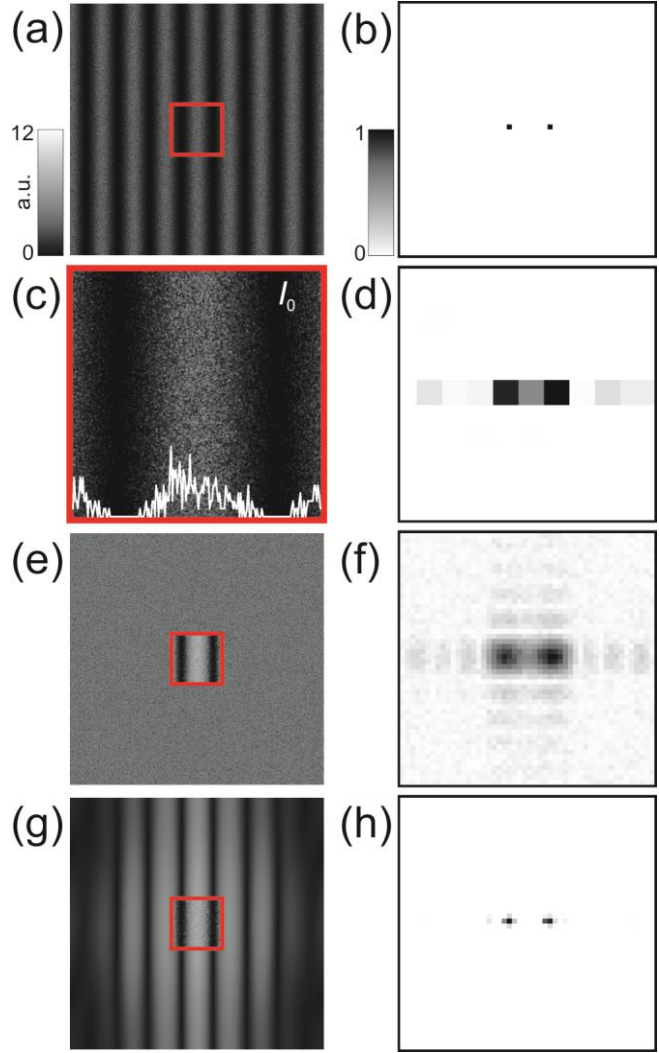

*Figure 1. Self-extrapolated interference pattern of two point sources. (a) Interference pattern created by two coherent point sources. (b) Amplitude of the transmission function of the two point sources. (c) $I_0$, a 200×200 pixels fragment of the interference pattern shown in (a) in the red square. (d) Object reconstructed from the interference pattern $I_0$ obtained after 500 iterations using hybrid input output algorithm[14] followed by 50 iterations using error reduction algorithm. The two point sources are not resolved. (e) Amplitude distribution obtained by padding $U_0$ with random complex-valued numbers. (f) Reconstruction of (e) after the first iteration. The two point sources are barely resolved. (g) Amplitude distribution of the self-extrapolated up to 1000×1000 pixels interference pattern after 10000 iterations using error-reduction algorithm. (h) Reconstruction of the self-extrapolated interference pattern shown in (g). The two point sources are clearly resolved.*

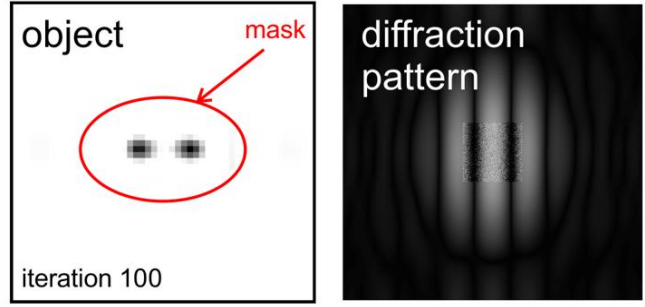

*Figure 2. Self-extrapolation of the far-field diffraction pattern created by two point scatterers after 100th iteration is shown.*

In the next example we simulated diffraction pattern of a real-valued object in form of a ""Ψ""-letter with four circles, an array of numbers decreasing in size for an easy visual inspection of resolution, and an array of bars for a quantitative measure of resolution, see Fig. 3(a). The diffraction pattern was sampled with 1000×1000 pixels, see Fig. 3(b), and at each pixel $I$(i,j) noise was added as a random value of a Gaussian distribution with mean $I$(i,j) and standard deviation $I$(i,j)/SNR, where SNR is the signal-to-noise ratio. Diffraction patterns without noise (SNR=∞), SNR=5 and SNR=2 were studied, see Fig.3(c). A fraction of interference pattern $I_0$ was obtained by cropping the simulated diffraction pattern to 500×500 pixels, as shown in Fig. 3(b). The post-extrapolation was performed as following. 20 reconstructions were obtained from $I_0$ using hybrid input output (HIO) algorithm[14], which was stopped after an object reconstruction and a local minimum in the error function were achieved. The results of these 20 reconstructions were averaged and the obtained complex-valued far-field distribution $U_0$ was used for extrapolation up to 1000×1000 pixels. The extrapolation was done using the error reduction (ER) algorithm[14] for 1000 iterations. The high-resolution part not available in $I_0$ (shown in the green square in Fig. 3(b)) was recovered and is qualitatively in good agreement with the related part of the original diffraction pattern, see Fig. 3(d). The object, reconstructed from the extrapolated diffraction pattern demonstrates superior resolution than the object reconstructed from the cropped diffraction pattern $I_0$, see Fig. 3(e) and (f). The width of the pale green bar in Fig. 3(f) equals to the Abbe limit $R_0=\lambda/(2N.A._0)$, and it is twice as large as the smallest distance between the object bars. As expected, the smallest distance between the object bars is not resolved in the reconstruction of the cropped diffraction pattern $I_0$, but it is well resolved in the reconstruction of the post-extrapolated diffraction pattern, even at SNR=2, see Fig.3(f). Thus, reconstruction beyond the Abbe' limit can be achieved even at low SNR.

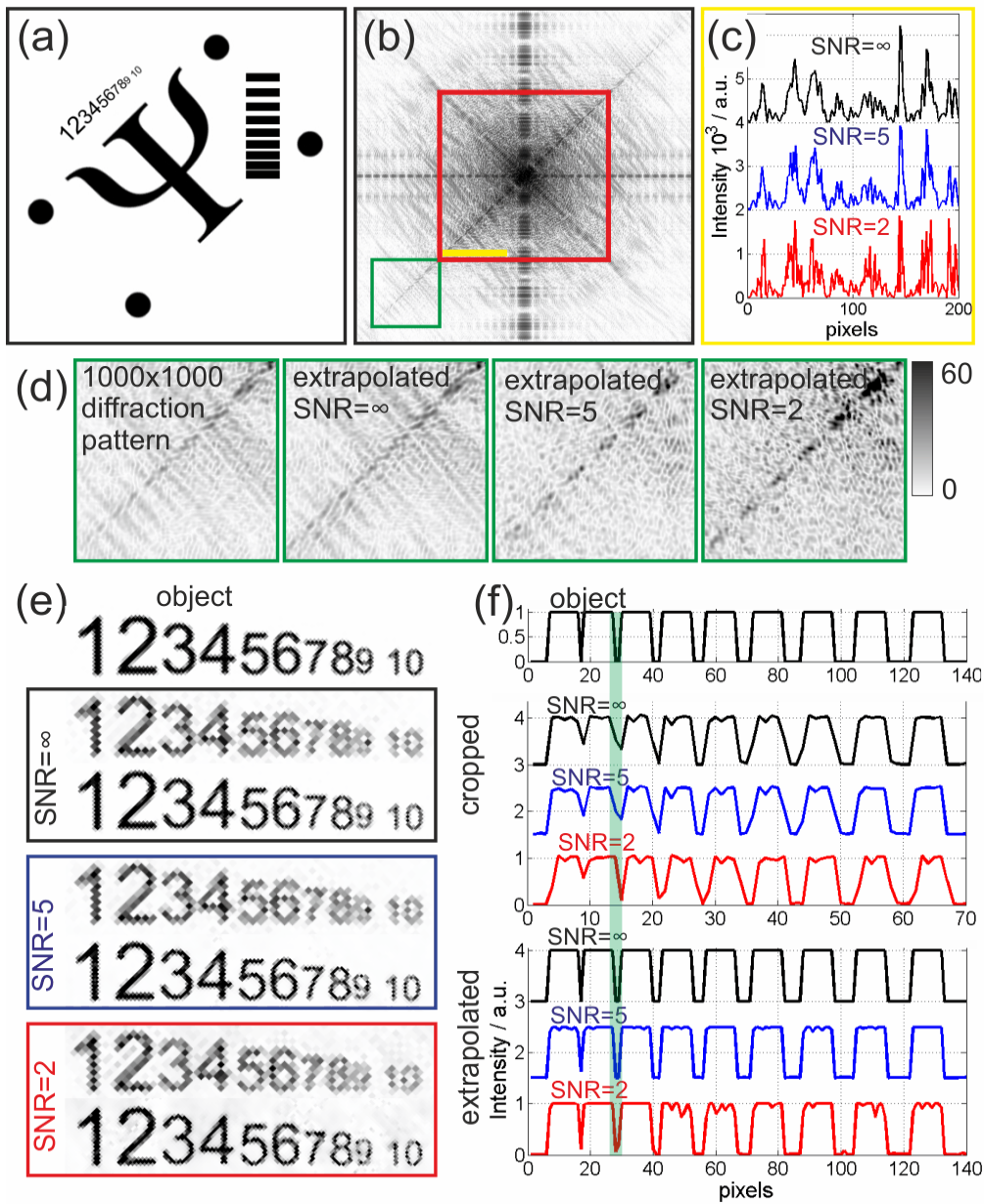

Figure 3. Simulated diffraction pattern of a real-valued object and its reconstruction by post-extrapolation. (a) A 400×400 pixels fragment of the object distribution; the total object area amounts to 1000×1000 pixels and thus the oversampling ratio amounts to σ=2.5. (b) Simulated diffraction pattern shown in logarithmic and inverted intensity scale. The region of 500×500 pixels, $I_0$, shown in the red square is used for the extrapolation procedure. (c) Intensity profiles along the yellow line in (b) at different SNR. (d) Magnified fragment of the original diffraction pattern, shown in (b) in the green square and the same region in the extrapolated diffraction pattern. (e) Original object and its reconstruction, selected part with the numbers is shown. In each pair, reconstructions from $I_0$ (top) and from the post-extrapolated diffraction pattern (bottom) are shown. (f) Profiles of the object bars in the original and reconstructed object distribution from $I_0$ (top) and the post-extrapolated diffraction pattern (bottom).

When it comes to experimental data the challenge is to correctly sample the waves constituting the interference pattern. Next to the Shannon-Nyquist sampling criterion[18-20], an accurate sampling of the *intensities* must also be fulfilled. In CDI, the intensity at the centre and the rim of diffraction pattern can differ by a few orders of magnitude. Conventionally, either a detector with intensity range of 16bit or higher is employed or a set of diffraction patterns at different exposure times is combined into a high-dynamic range image.

Optical diffraction patterns were recorded using 532 nm wavelength laser light. As sample we used a microscope cover slip on which a thin layer of gold was evaporated. A focussed ion beam was used to engrave a pattern displaying a "Ψ", a "2" and four circles, as shown in Fig. 4(a). The diffraction pattern of this sample, recorded with an oversampling ratio of σ=9 in both dimensions, is shown in Fig. 4(b), and its reconstruction in Fig. 4(c). The intrinsic resolution of the recorded diffraction pattern[7], according to the Abbe criterion, amounts to 1.8 μm, being in good agreement with the quality of the reconstruction, shown in Fig. 4(c). Next, we crop the diffraction pattern keeping only its central 300×300 pixels part $I_0$, as depicted by the red square in Fig. 4(b) and also in Fig. 4(d). The intrinsic resolution of the cropped diffraction pattern $I_0$ amounts to only $R_0=\lambda/(2N.A._0)=5.9$ μm. As a consequence, its reconstruction resembles a blurred object, shown in Fig. 4(e). Then we apply our extrapolation technique to recreate the high-resolution information from the cropped diffraction pattern $I_0$. The complex-valued field distribution $U_0$ at the detector is padded up to 1000×1000 pixels with random complex-valued numbers and reconstructed with the same algorithm as already described above. After 1000 iterations the diffraction pattern has extrapolated itself beyond $I_0$ as shown in Fig. 4(f). Due to this effectively increased N.A, the reconstruction of the self-extrapolated diffraction pattern, shown in Fig. 4(g), exhibits an enhanced resolution compared to Fig. 4(e), fine fringes are now apparent and well resolved. Quantitatively, resolution beyond the Abbe limit is achieved: $R=\lambda/(2N.A.)=1.8$ μm. Thus, the post-experimental treatment of the detected wave field allows circumventing the resolution limit imposed by the Abbe criterion. To cross-validate our method, we also performed the same iterative reconstruction procedure but with $I_0$ zero padded[21-22] during the entire retrieval routine, as shown in Fig. 4(h). The result is just a blurred reconstruction of the original object, see Fig. 4(i).

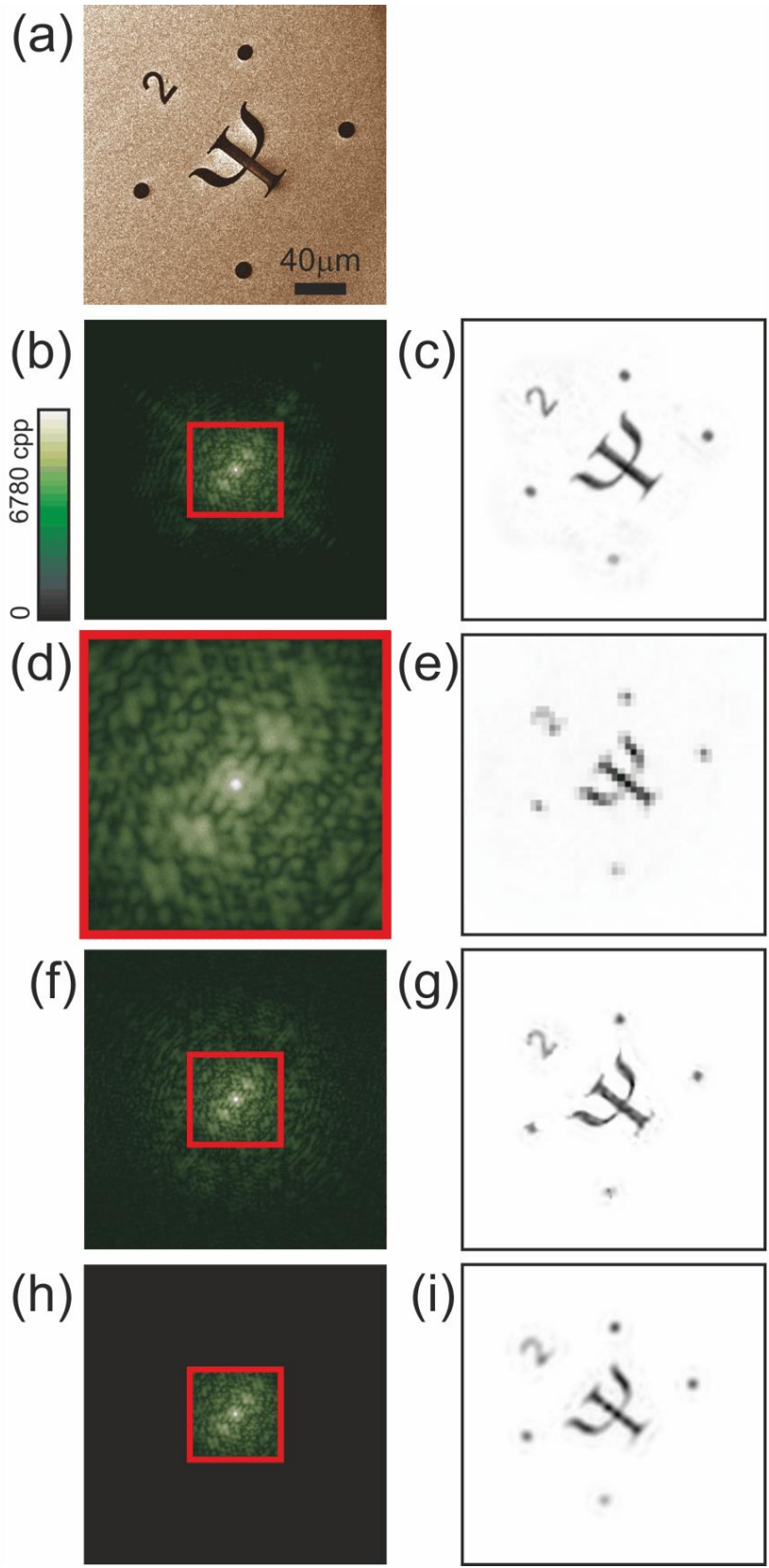

*Figure 4. Self-extrapolated experimental diffraction pattern. (a) Scanning electron microscope image of the sample. (b) Amplitude of the optical diffraction pattern, recorded at 60 cm distance from the sample with 532 nm wavelength laser light. The diffraction pattern was imaged at a screen of size*

*181×181 mm² and sampled with 1000×1000 pixels using 10 bit camera. To capture the intensity variations in the central part and at the rim of the diffraction pattern, a set of images was recorded at various exposures and those were combined into a high dynamic range diffraction pattern. The resulting diffraction pattern displays intensities up to 6780 counts per pixel (cpp). (c) Its reconstruction. After 64 iteration using HIO algorithm with the feedback parameter β = 0.9 and a loose mask support (with a diameter 2.3 times larger than that of the object) in the object domain, the mask was updated to a more tight one (with a diameter 2 times larger than that of the object) and additional 1000 iterations using ER algorithm were performed. (d) A 300×300 pixels fragment of the amplitude of the experimental diffraction $I_0$, indicated with red square. (e) Reconstruction of the fragment of the diffraction pattern $I_0$ resulting from averaging over 50 reconstructions obtained with HIO algorithm. (f) Amplitude of the self-extrapolated up to 1000×1000 pixels diffraction pattern after total 1000 iterations using ER algorithm. (g) Reconstruction of the self-extrapolated diffraction pattern shown in (f). (h) Amplitude of the zero-padded diffraction pattern. (i) Its reconstruction achieved after 1000 iterations using ER algorithm.  Amplitude of the diffraction pattern in (b), (d), (f) and (h) is shown in logarithmic scale.*

We have demonstrated that even an incomplete fraction of an interference pattern contains already enough information to extrapolate the wave field far beyond the actual data collected. Our method is applicable to any interference patterns created by elastic scattering from a non-periodic object. A limited size low-resolution interference pattern is sufficient to recreate a high-resolution reconstruction of the object. This implies that, even without any additional experiments, the resolution in previously reconstructed experimental data can be post-enhanced by applying our technique. While our technique can be applied to any kind of radiation, be it be X-rays, photons or electrons, the following conditions must be fulfilled: (1) A sufficiently coherent source must be used in order to provide an interference pattern with good contrast. (2) The detector should be capable to capture the interference pattern with a high dynamic intensity range. (3) The scattering object must be of finite size. Although we related this tool to diffraction patterns here, the method can be applied to any other interference patterns, for instance, created by Fresnel coherent diffraction imaging[23], Fourier-transform[24] or classical Gabor type holography[4-5].

## Acknowledgements

We would like to thank Michael Krueger for the sample preparation. The work presented here is financially supported by the Swiss National Science Foundation (SNF).